\documentclass[10pt,aps,prd,twocolumn,superscriptaddress,showpacs,nofootinbib,showkeys]{revtex4-1}
\pdfoutput=1

\usepackage{graphicx}
\usepackage{subfigure}
\usepackage{bm} 

\begin{document}
 
\title{Chromoelectric oscillations in a dynamically evolving anisotropic background}

\author{Wojciech Florkowski} 
\affiliation{Institute of Physics, Jan Kochanowski University, PL-25406~Kielce, Poland} 
\affiliation{The H. Niewodnicza\'nski Institute of Nuclear Physics, Polish Academy of Sciences, PL-31342 Krak\'ow, Poland}

\author{Radoslaw Ryblewski} 
\affiliation{The H. Niewodnicza\'nski Institute of Nuclear Physics, Polish Academy of Sciences, PL-31342 Krak\'ow, Poland} 

\author{Michael Strickland} 
\affiliation{Physics Department, Gettysburg College\\
Gettysburg, PA 17325 United States}
\affiliation{Frankfurt Institute for Advanced Studies\\
Ruth-Moufang-Strasse 1\\
D-60438, Frankfurt am Main, Germany}

\date{\today}

\begin{abstract}
We study the oscillations of a uniform longitudinal chromoelectric field in a dynamically-evolving momentum-space anisotropic 
background in the weak field limit.  Evolution equations for the background are derived by taking moments of the Boltzmann
equation in two cases: (i) a fixed relaxation time and (ii) a relaxation time that is proportional to the local inverse transverse 
momentum scale of the plasma.  The second case allows us to reproduce 2nd-order viscous hydrodynamical dynamics in the 
limit of small shear viscosity to entropy ratio.  We then linearize the Boltzmann-Vlasov equation in a dynamically-evolving 
background and obtain an integro-differential evolution equation for the chromoelectric field.  We present numerical solutions 
to this integro-differential equation for a variety of different initial conditions and shear viscosity to entropy density ratios.  The 
dynamical equations obtained are novel in that they include a non-trivial time-dependent momentum-space anisotropic 
background and the effect of collisional damping for the first time.
\end{abstract}



\maketitle 

\section{Introduction}
\label{sect:intro}

The purpose of ongoing and upcoming heavy ion collision experiments at the Relativistic Heavy Ion Collider (RHIC) and the Large 
Hadron Collider (LHC) is to study the behavior of nuclear matter at high energy density, $\epsilon \gg 1\;{\rm GeV/fm}^3$.
At such high energy densities one expects to create a deconfined quark gluon plasma (QGP).  With such experiments one hopes 
to not only cross the threshold necessary to create a QGP, but to also study its properties such as transport coefficients, color 
opacity, etc.  One complicating factor is that the QGP generated in such collisions lasts for only a few fm/c and during this time 
the bulk properties of the system, e.g. energy density and pressure, can change rapidly.  Therefore, dynamical models that can 
describe the evolution of the system on the fm/c timescale are necessary in order to make reliable phenomenological 
predictions.

To first approximation, it seems that the dynamics of the soft background is  well-described by relativistic viscous 
hydrodynamics \cite{Israel:1979wp,Muronga:2003ta,Baier:2006um,Romatschke:2007mq,Dusling:2007gi,Luzum:2008cw,%
Song:2008hj,Denicol:2010tr,Schenke:2011tv,Shen:2011eg,Bozek:2011wa,Niemi:2011ix,Bozek:2012qs}.   However, 
viscous corrections to the ideal energy momentum tensor cause it to become anisotropic in the local rest frame of the system.  
For small deviations from isotropy, 2nd-order viscous hydrodynamics describes the evolution
quite well; however, for large deviations from isotropy this is no longer the case.  Large deviations from isotropy occur
at very early times after the initial nuclear impact and near the transverse or longitudinal edges of the plasma where the
matter is nearly free streaming.  The presence of momentum-space anisotropies seems unavoidable in dynamical 
models.  In fact, even in the limit of infinite strong coupling, momentum-space anisotropies persist during the entire 
lifetime of the plasma  \cite{Heller:2011ju,Heller:2012je,Wu:2011yd}.  Large momentum-space anisotropies pose a 
problem for 2nd-order viscous hydrodynamics since it relies on a linearization around an isotropic
background.  If the linear corrections grow too large this can generate unphysical results such as negative particle 
pressures, negative one-particle distribution functions, etc.~\cite{Martinez:2009mf}.

In order to ameliorate these problems it is possible to reorganize the derivation of the necessary dynamical equations by 
linearizing around an anisotropic instead of isotropic background.  Doing so results in a dynamical framework called anisotropic 
hydrodynamics \cite{Florkowski:2010cf,Martinez:2010sc,Ryblewski:2010bs,Martinez:2010sd,Ryblewski:2011aq,
Martinez:2012tu,Ryblewski:2012rr}.  In the limit of small deviations from isotropy, anisotropic hydrodynamics reduces
to 2nd-order viscous hydrodynamics, but can also faithfully describe large deviations from isotropy such as those
created during the initial longitudinal free streaming phase of the plasma lifetime.  This framework has now been
used to model the full (3+1)-dimensional dynamics of the QGP \cite{Ryblewski:2012rr}.  Comparison of the anisotropic
hydrodynamics predictions for observables such as the bulk flow as a function of transverse momentum and 
rapidity with experimental data indicate that it is possible that large momentum-space anisotropies can persist for 
up to 2 -- 3 fm/c after the initial nuclear impact.  Given this, it is imperative to revisit the study of basic properties
of the QGP in a time-evolving anisotropic background.

In this paper we study the oscillations of a uniform longitudinal chromoelectric field in a dynamically-evolving 
momentum-space anisotropic background in the weak field limit.  For simplicity, in this work we restrict ourselves 
to a (0+1)-dimensional boost-invariant background.  The necessary anisotropic hydrodynamics equations are
obtained from the first two moments of the Boltzmann-Vlasov equation using a spheroidal form for the one-particle
distribution function in the local rest frame \cite{Romatschke:2003ms}.  The dynamical equations in this case
were first obtained in Refs.~\cite{Florkowski:2010cf,Martinez:2010sc}.  In both Ref.~\cite{Florkowski:2010cf}
and Ref.~\cite{Martinez:2010sc} a timescale for the 
approach to isotropic thermal equilibrium, $\tau_{\rm eq}$, was introduced.  In Ref.~\cite{Florkowski:2010cf}
$\tau_{\rm eq}$ was assumed to be a constant, while in Ref.~\cite{Martinez:2010sc} this time scale
was proportional to the average local inverse transverse momentum of the plasma constituents and was determined
self-consistently in terms of the local plasma environment.

In the case that $\tau_{\rm eq}$ is constant, the late time behavior of the system is that of ideal hydrodynamics.  
In the case that $\tau_{\rm eq}$ is proportional to the local inverse transverse momentum scale, the 
proportionality constant can be fixed by requiring that the late time dynamics of the system is that of 2nd-order 
viscous hydrodynamics.  We will consider both cases in order to assess the impact of this choice.  In either case the
anisotropic hydrodynamics equations provide the proper-time dependence of the local transverse momentum scale,
$\Lambda(\tau)$, and momentum-space anisotropy $\xi(\tau) = \frac{1}{2} \langle p_\perp^2 \rangle/\langle 
p_\parallel^2 \rangle - 1$ where $\langle p_\perp^2 \rangle$ and $\langle p_\parallel^2 \rangle$ are the average transverse 
and longitudinal (beamline-direction) momenta squared in the local rest frame of the plasma constituents,
respectively.

Given this time evolving background, we linearize the Boltzmann-Vlasov equation in order to study the evolution
of a uniform longitudinal chromoelectric  field fluctuation, $\mbox{\boldmath $\cal E$}_z$.  
We consider the weak-field limit in which case
we can use the abelian dominance approximation for the color fields \cite{Gyulassy:1986jq,Elze:1986hq,%
Elze:1986qd,Bialas:1987en} (see also Ref.~\cite{Casher:1978wy}).  In a static constant-temperature plasma, 
uniform longitudinal field fluctuations 
oscillate in time with a frequency given by the plasma frequency $\omega_{\rm pl} = m_D/\sqrt{3}$ where 
$m_D^2 = (N_c/3+ N_f/6) g^2 T^2$ is the leading-order gluonic Debye mass.  In a time evolving system, the plasma 
frequency is time dependent and one must self-consistently solve the linearized Boltzmann-Vlasov equation together with 
the Maxwell equations.  In general, the result can be cast in the form of an integro-differential equation for the 
evolution of $\mbox{\boldmath $\cal E$}_z$.  
For the case of ideal hydrodynamical evolution, Bialas and Czyz \cite{Bialas:1987cn} derived 
such an equation and solved it numerically.  

Here we extend the treatment of Bialas and Czyz to (i) include a dynamically evolving anisotropic background and 
(ii) include the effect of collisional damping.  We will present numerical solutions to the resulting 
integro-differential equations for small and large magnitude momentum-space anisotropies in order to assess the 
impact of momentum-space anisotropy on plasma oscillations.  The equations we obtain are applicable to an arbitrary
time-dependent anisotropic background.  Although we consider the evolution of a stable longitudinal chromoelectric 
field, the techniques used herein could have application to the study of the evolution of non-Abelian plasma
instabilities in a dynamically evolving anisotropic background \cite{Romatschke:2006wg,Rebhan:2008uj,Rebhan:2009ku}.
As a cross check of our results we present a comparison with the results of Ref.~\cite{Rebhan:2009ku} which
presented an analysis of all stable and unstable collective modes of the QGP in the limit of a longitudinally free streaming
background.  We show that in this limit we obtain the same evolution and asymptotic behavior as in 
Ref.~\cite{Rebhan:2009ku}, giving us confidence in our theoretical and numerical methods.

The structure of the paper is as follows.  In Section \ref{sect:conventions} we specify the conventions we will
use throughout the paper.  In Section \ref{sect:semiclassical} we review the semi-classical transport equations
for the quark gluon plasma in the abelian dominance approximation.  In Section \ref{sect:linearization} we 
linearize the Boltzmann-Vlasov equation to zeroth and first order in fluctuations.  In Section \ref{sect:Max}
we couple the fluctuations via currents to the Maxwell equations and solve the coupled Boltzmann-Vlasov-Maxwell
system of equations to obtain an integro-differential equation which governs the time evolution of uniform longitudinal
chromoelectric fluctuations.  In Section \ref{sect:results} we present the results of numerical solution to the
evolution equations for different types of anisotropic backgrounds and compare with numerical and analytic results 
available in the limit of longitudinal free streaming.  In Section \ref{sect:concl} we present our conclusions and an 
outlook for the future.

\section{Conventions}
\label{sect:conventions}

Below we use the following definitions for momentum rapidity ($y$) and spacetime rapidity ($\eta$),
\begin{eqnarray}
y = \frac{1}{2} \ln \frac{E+p_\parallel}{E-p_\parallel}, \quad
\eta = \frac{1}{2} \ln \frac{t+z}{t-z}, \label{yandeta} 
\end{eqnarray}
which come from the standard parameterization of the four-momentum and spacetime coordinates of a particle,
\begin{eqnarray}
p^\mu &=& \left(E, {\vec p}_\perp, p_\parallel \right) =
\left(m_\perp \cosh y, {\vec p}_\perp, m_\perp \sinh y \right), \nonumber \\
x^\mu &=& \left( t, {\vec x}_\perp, z \right) =
\left(\tau \cosh \eta, {\vec x}_\perp, \tau \sinh \eta \right). \label{pandx}
\end{eqnarray} 
In Eq.~(\ref{pandx}) the quantity $m_\perp$ is the transverse mass
\begin{equation}
m_\perp = \sqrt{m^2 + p_x^2 + p_y^2},
\label{energy}
\end{equation}
and $\tau$ is the proper time
\begin{equation}
\tau = \sqrt{t^2 - z^2}.
\label{tau}
\end{equation} 
Throughout the paper we use natural units where $c=1$ and $\hbar=1$.

\section{Semi-classical kinetic equations for quark-gluon plasma}
\label{sect:semiclassical}

In the abelian dominance approximation, the transport equations for quarks, antiquarks, and gluons have the form 
\cite{Gyulassy:1986jq,Elze:1986hq,Elze:1986qd,Bialas:1987en} 
\begin{equation}
\left( p^{\mu }\partial _{\mu } \pm g{\mbox{\boldmath $\epsilon$}}_{i}\cdot 
{\bf F}^{\mu \nu }p_{\nu }\partial _{\mu }^{p}\right) 
Q^\pm_{i}(x,p)= C^\pm_{i},  \label{kineq}
\end{equation}
\begin{equation}
\left( p^{\mu }\partial _{\mu }+g{\mbox{\boldmath $\eta$}}_{ij}\cdot 
{\bf F}^{\mu \nu }p_{\nu }\partial _{\mu }^{p}\right) 
G_{ij}(x,p)=
C_{ij},  \label{kineg}
\end{equation}
where $Q^+_{i}(x,p)$, $Q^-_{i}(x,p)$, and $G_{ij}(x,p)$ are the phase-space densities of quarks, antiquarks, and charged 
gluons, respectively. Here $g$ is the strong coupling constant, and $i,j=(1,2,3)$ are color indices. The terms on the 
left-hand-side describe the free motion of the particles and the interaction of the particles with the mean field 
$\mathbf{F}_{\mu \nu }$.   The latter describes neutral gluons \cite{Huang:1982ik}.  

In this work, the only non-zero components of the tensor ${\bf F}^{\mu \nu }=(F^{\mu \nu }_{(3)},F^{\mu \nu }_{(8)})
$ are those corresponding to the longitudinal chromoelectric  field ${\mbox{\boldmath $\cal E$}_z} = 
(F^{30}_{(3)},F^{30}_{(8)})$. The quarks couple to the chromoelectric field ${\mbox{\boldmath $\cal E$}_z}$ through the 
charges 
\begin{equation}
\mbox{\boldmath $\epsilon$}_{1} = \frac{1}{2}\left(\! 1,\sqrt{\frac{1}{3}}\right) \!, 
\mbox{\boldmath $\epsilon$}_{2} = \frac{1}{2}\left(\! -1,\sqrt{\frac{1}{3}}\right) \!, 
\mbox{\boldmath $\epsilon$}_{3} = \left(\! 0,-\sqrt{\frac{1}{3}}\right) .
\label{qcharge}
\end{equation}
The gluons couple to ${\mbox{\boldmath $\cal E$}_z}$ through the charges ${\mbox{\boldmath $\eta $}}_{ij}$ defined by 
the relation 
\begin{equation}
{\mbox{\boldmath $\eta$}}_{ij}={\mbox{\boldmath $\epsilon$}}_{i}-{%
\mbox{\boldmath $\epsilon$}}_{j}.  \label{gcharge}
\end{equation}
Below, we make use of the following relations
\begin{eqnarray}
\sum_{i=1}^3 {\mbox{\boldmath $\epsilon$}}^a_{i} {\mbox{\boldmath $\epsilon$}}^b_{i} 
&=& \frac{1}{2} \, \delta^{ab} \, , \nonumber \\
\sum_{i,j=1}^3 {\mbox{\boldmath $\eta$}}^a_{ij} {\mbox{\boldmath $\eta$}}^b_{ij} 
&=& 3 \, \delta^{ab} \, , \label{norm}
\end{eqnarray}
where $a,b  \in \{ 3,8 \}$.

The terms on the right-hand-side of Eqs.~(\ref{kineq}) and (\ref{kineg}) are collisions terms, which we treat in the relaxation 
time approximation
\begin{eqnarray}
C^\pm_{i} &=& - p^\mu U_\mu \frac{Q^\pm_{i}(x,p)-Q^\pm_{\rm eq}(x,p)}{\tau_{\rm eq}}, \\
C_{ij} &=& - p^\mu U_\mu \frac{G_{ij}(x,p)-G_{\rm eq}(x,p)}{\tau_{\rm eq}}.
\end{eqnarray}
Here $U^\mu$ is the four-velocity of the local rest frame 
\begin{equation}
U^\mu = \gamma (1, v_x, v_y, v_z), \gamma=(1-v^2)^{-1}.
\end{equation}
In this paper we consider boost-invariant longitudinal expansion and hence we set $v_x=v_y=0$ and $v_z=z/t$.

\section{Linearization of kinetic equations}
\label{sect:linearization}

In the following, we seek the solutions of kinetic equations of the form
\begin{eqnarray}
Q^\pm_{i}(x,p) = Q^\pm_{0}(x,p) + \delta Q^\pm_{i}(x,p), \\
G_{ij}(x,p) = G_{0}(x,p) + \delta G_{ij}(x,p),
\end{eqnarray}
where the corrections to the background distributions are proportional to the coupling. We emphasize that the background 
distributions $Q^\pm_{0}(x,p)$ and $G_{0}(x,p)$ are different from the equilibrium distributions. 

\subsection{Zeroth order}

At zeroth order in fluctuations one obtains 
\begin{equation}
 p^{\mu }\partial_{\mu } Q^\pm_{0}(x,p)= 
- p^\mu U_\mu \frac{Q^\pm_{0}(x,p) - Q^\pm_{\rm eq}(x,p)}{\tau_{\rm eq}},  
\label{kineq0}
\end{equation}
\begin{equation}
p^{\mu }\partial_{\mu } G_{0}(x,p) = 
- p^\mu U_\mu \frac{G_{0}(x,p) - G_{\rm eq}(x,p)}{\tau_{\rm eq}}.  
\label{kineg0}
\end{equation} 
Equations (\ref{kineq0}) and (\ref{kineg0}) determine the evolution of $Q^\pm_{0}(x,p)$ and $G_{0}(x,p)$. 

Instead of solving (\ref{kineq0}) and (\ref{kineg0}) directly, we take moments of these equations.  
In order to describe (0+1)-dimensional anisotropic dynamics we take the zeroth and first moments of 
Eqs.~(\ref{kineq0}) and (\ref{kineg0}) assuming that the distributions $Q^\pm_{0}(x,p)$ and 
$G_{0}(x,p)$ are given by the covariant version of the 
Romatschke-Strickland distribution \cite{Romatschke:2003ms,Florkowski:2011jg}, namely
\begin{eqnarray}
Q^\pm_{0}(x,p) = G_{0}(x,p) = f_0(x,p) ,
\label{f0}
\end{eqnarray}
where
\begin{eqnarray}
f_0(x,p) = \exp\left(-\frac{1}{\Lambda} \sqrt{(p\cdot U)^2 + \xi (p\cdot V)^2} \right).
\label{RSform}
\end{eqnarray}
Accordingly, we take
\begin{eqnarray}
Q^\pm_{\rm eq}(x,p) = G_{\rm eq}(x,p) = f_{\rm eq}(x,p),
\label{feq}
\end{eqnarray}
where
\begin{eqnarray}
f_{\rm eq}(x,p) = \exp\left(-\frac{p\cdot U}{T}  \right).
\label{eqform}
\end{eqnarray}
Note that one can also use anisotropic Fermi-Dirac and Bose-Einstein distributions for the (anti-)quarks and
gluons, respectively; however, the only change to the final result will be the precise value of the isotropic
plasma frequency of the system.  For the sake of simplicity we present the case of a Boltzmann distribution
and generalize the results to the full quantum statistical distributions in the end.

The four-vector $V^\mu$ appearing in (\ref{RSform}) defines the direction of the beam ($z$-axis)
\begin{equation}
V^\mu = \gamma_z (v_z, 0, 0, 1), \quad \gamma_z = (1-v_z^2)^{-1/2}.
\label{V}
\end{equation}
We note that the four-vectors $U^\mu$ and $V^\mu$ satisfy the normalization conditions
\begin{eqnarray}
U^2 = 1, \quad V^2 = -1, \quad U \cdot V = 0.
\label{UVnorm}
\end{eqnarray}
In the local rest frame of the fluid element, $U^\mu$ and $V^\mu$ have simple forms
\begin{eqnarray}
 U^\mu = (1,0,0,0), \quad V^\mu = (0,0,0,1). 
 \label{UVLRF}
\end{eqnarray}
For the (0+1)-dimensional boost-invariant expansion considered in this paper, we may use
\begin{eqnarray}
 U^\mu &=& (\cosh\eta,0,0,\sinh\eta), \nonumber \\
 V^\mu &=& (\sinh\eta,0,0,\cosh\eta).
 \label{UVbinv}
\end{eqnarray}

With the assumptions (\ref{f0}) and (\ref{feq}), the kinetic equations (\ref{kineq0}) and (\ref{kineg0}) are reduced to a
single equation for the background distribution
\begin{equation}
 p^{\mu }\partial_{\mu } f_{0}(x,p)= 
- p^\mu U_\mu \frac{f_{0}(x,p) - f_{\rm eq}(x,p)}{\tau_{\rm eq}}.
\label{kinef0}
\end{equation}

\subsubsection{Zeroth moment of the kinetic equation}

Integrating Eq.~(\ref{kinef0}) over three-momentum and including the internal degrees of freedom we obtain
\begin{equation}
\partial_{\mu } N_{0}^\mu = 
 \frac{U_\mu \left( N_{\rm eq}^\mu-N_{0}^\mu \right)}{\tau_{\rm eq}},
\label{EQN}
\end{equation}
where $N_0$ and $N_{\rm eq}$ are particle currents~\footnote{There is no term proportional to $V^\mu$ in $N_0^\mu$, due 
to the quadratic dependence of $f_0$ on $V^\mu$.} 
\begin{equation}
N_0^\mu = n_0 U^\mu, \quad N^\mu_{\rm eq} = n_{\rm eq} U^\mu.
\label{N0Neq}
\end{equation}
A simple calculation performed in the local rest frame gives
\begin{eqnarray}
n_0 = \frac{g_0}{\pi^2} \frac{\Lambda^3}{\sqrt{1+\xi}}, \quad
n_{\rm eq} = \frac{g_0}{\pi^2} T^3.
\label{n0neq}
\end{eqnarray}
Here $g_0$ is the degeneracy factor accounting for internal degrees of freedom (we show below that the equations of motion for 
the background are insensitive to the specific choice of $g_0$). For longitudinal boost-invariant expansion one finds
\begin{equation}
U^\mu \partial_\mu = \frac{d}{d\tau}, \quad \partial_\mu U^\mu = \frac{1}{\tau}.
\end{equation}
Thus, using (\ref{N0Neq}) and (\ref{n0neq}) in (\ref{EQN}), we obtain 
\begin{eqnarray}
\frac{3}{\Lambda} \frac{d\Lambda}{d\tau}-\frac{1}{2(1+\xi)}\frac{d\xi}{d\tau} + \frac{1}{\tau} 
=   \frac{\left(T/\Lambda\right)^3  \sqrt{1+\xi} -1}{\tau_{\rm eq}}.
\label{EQ1}
\end{eqnarray}

\subsubsection{First moment of the kinetic equation}

In the next step we multiply Eq.~(\ref{kinef0}) by $p^\nu$ and integrate over three-momentum. In this way, we obtain
\begin{eqnarray}
\partial_\mu T_0^{\mu \nu} = 
\frac{U_\mu \left(T_{\rm eq}^{\mu \nu}-T_0^{\mu \nu}\right)}{\tau_{\rm eq}},
\label{enmomcon0}
\end{eqnarray}
where \cite{Florkowski:2010cf,Martinez:2012tu}
\begin{equation}
T_0^{\mu \nu} = (\varepsilon_0+P_\perp) U^\mu U^\nu - P_\perp g^{\mu \nu}
-(P_\perp-P_\parallel) V^\mu V^\nu
\label{Tmunu0}
\end{equation}
and
\begin{equation}
T_{\rm eq}^{\mu \nu} = (\varepsilon_{\rm eq}+P_{\rm eq}) U^\mu U^\nu - P_{\rm eq} g^{\mu \nu}.
\label{Tmunu0-2}
\end{equation}

In order to conserve energy and momentum, the right-hand-side of (\ref{enmomcon0}) should vanish. Hence, we obtain 
the Landau matching condition 
\begin{equation}
\varepsilon_0 = \varepsilon_{\rm eq},
\label{LM1}
\end{equation}
where
\begin{eqnarray}
\varepsilon_0 = \frac{3 g_0 \Lambda^4}{\pi^2} {\cal R}(\xi), \quad
\varepsilon_{\rm eq} = \frac{3 g_0 T^4}{\pi^2},
\label{eps0epseq}
\end{eqnarray}
and the function ${\cal R}(\xi)$ has the form \cite{Martinez:2010sc}
\begin{equation}
{\cal R}(\xi) = \frac{1}{2(1+\xi)} 
\left[1+ \frac{ (1+\xi) \arctan \sqrt{\xi}} {\sqrt{\xi} } \right].
\end{equation}
Eqs.~(\ref{LM1}) and (\ref{eps0epseq}) are used to obtain the ratio $T/\Lambda$ needed in (\ref{EQ1})
\begin{equation}
T  = \Lambda {\cal R}^{1/4}(\xi).
\label{EQ2}
\end{equation} 

For purely longitudinal boost-invariant motion, the energy-momentum conservation law $\partial_\mu T_0^{\mu \nu} =0$ takes a simple form
\begin{equation}
\frac{d\varepsilon_0}{d\tau} = -\frac{\varepsilon_0+P_\parallel}{\tau}.
\label{EQ30}
\end{equation}
Eq.~(\ref{EQ30}) may be reduced to the equation
\begin{eqnarray}
\hspace{-4mm} {\cal R}^\prime (\xi) \frac{d\xi}{d\tau} + 4 {\cal R}(\xi) \frac{d\Lambda}{\Lambda d\tau}
= -\frac{1}{\tau} \left( {\cal R}(\xi) + \frac{1}{3} {\cal R}_L(\xi) \right) ,
\label{EQ3}
\end{eqnarray}
where
\begin{equation}
{\cal R}_L(\xi) = \frac{3}{\xi} \left[ {\cal R}(\xi)
- \frac{1}{1+\xi} \right].
\end{equation}

\subsubsection{Evolution of the time-evolving background}
\label{subsect:background}

Eqs.~(\ref{EQ1}), (\ref{EQ2}), and (\ref{EQ3}) provide three equations for three unknown functions:  $\Lambda(\tau)$, 
$\xi(\tau)$, and $T(\tau)$. The solutions of these equations allow us to determine the background for the plasma oscillations.  
Below we will consider two cases:  (i) a fixed relaxation time of $\tau_{\rm eq} = $~1~fm/c and (ii) a relaxation time that is 
proportional to the local inverse transverse momentum scale of the plasma.  In the second case the relaxation 
time is fixed by requiring that, in the limit of small momentum-anisotropy, the linearized anisotropic hydrodynamics 
equations reproduce 2nd-order viscous hydrodynamics \cite{Martinez:2010sc}.  In this case, the relaxation time is 
given by
\begin{equation}
\tau_{\rm eq}(\tau) = \frac{5 \bar\eta}{2 {\cal R}^{1/4}(\xi) \Lambda} \, ,
\label{taueqdyn}
\end{equation}
where $\bar\eta = \eta_s/{\cal S}$ is the ratio of the shear viscosity to entropy density and
it is implicitly understood that $\xi$ and $\Lambda$ depend on proper time.  In what
follows we will assume that $\bar\eta$ is time independent.

\begin{figure}[t]
\begin{center}
\subfigure{\includegraphics[angle=0,width=0.42\textwidth]{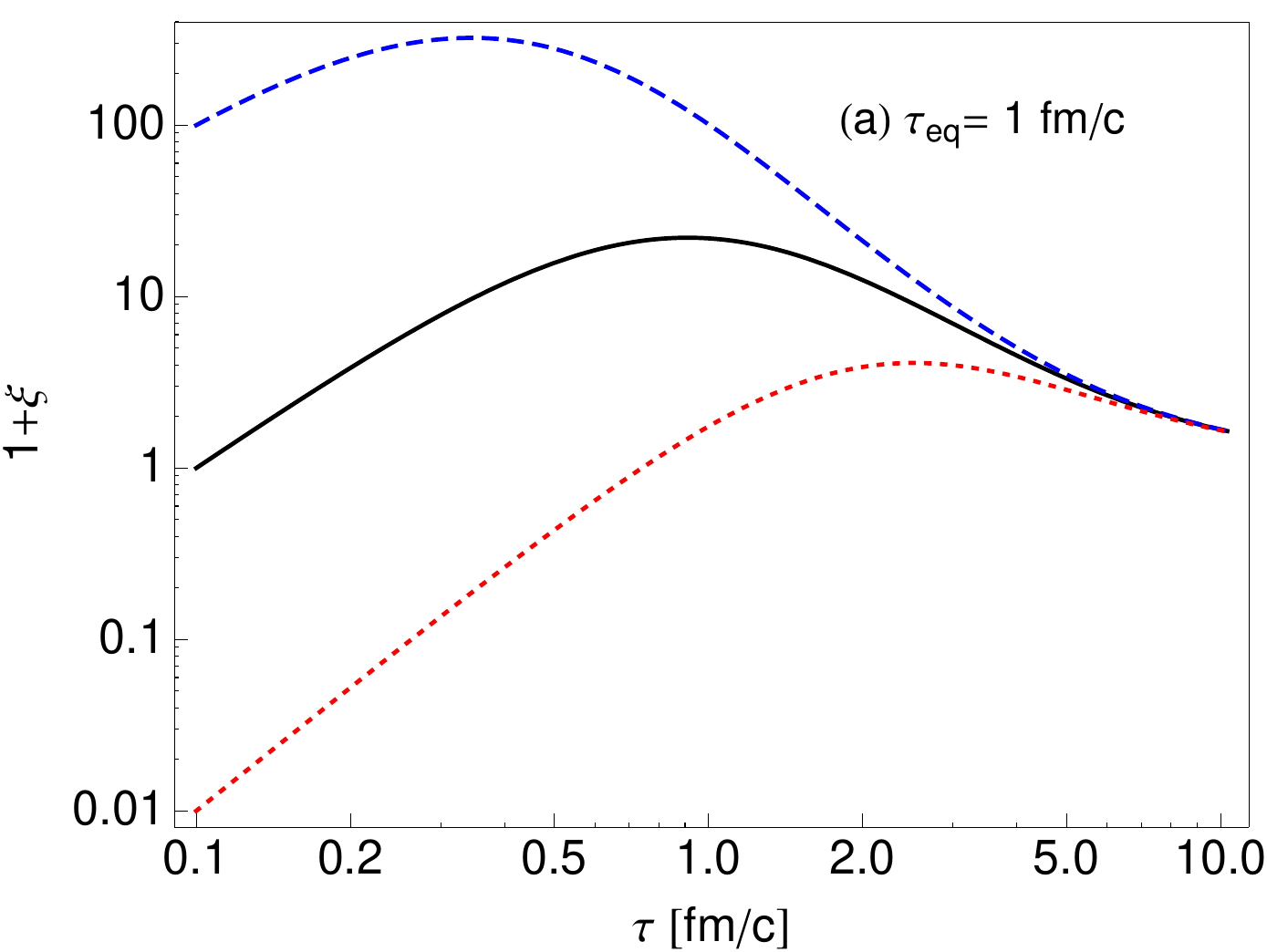}}\\
\subfigure{\includegraphics[angle=0,width=0.42\textwidth]{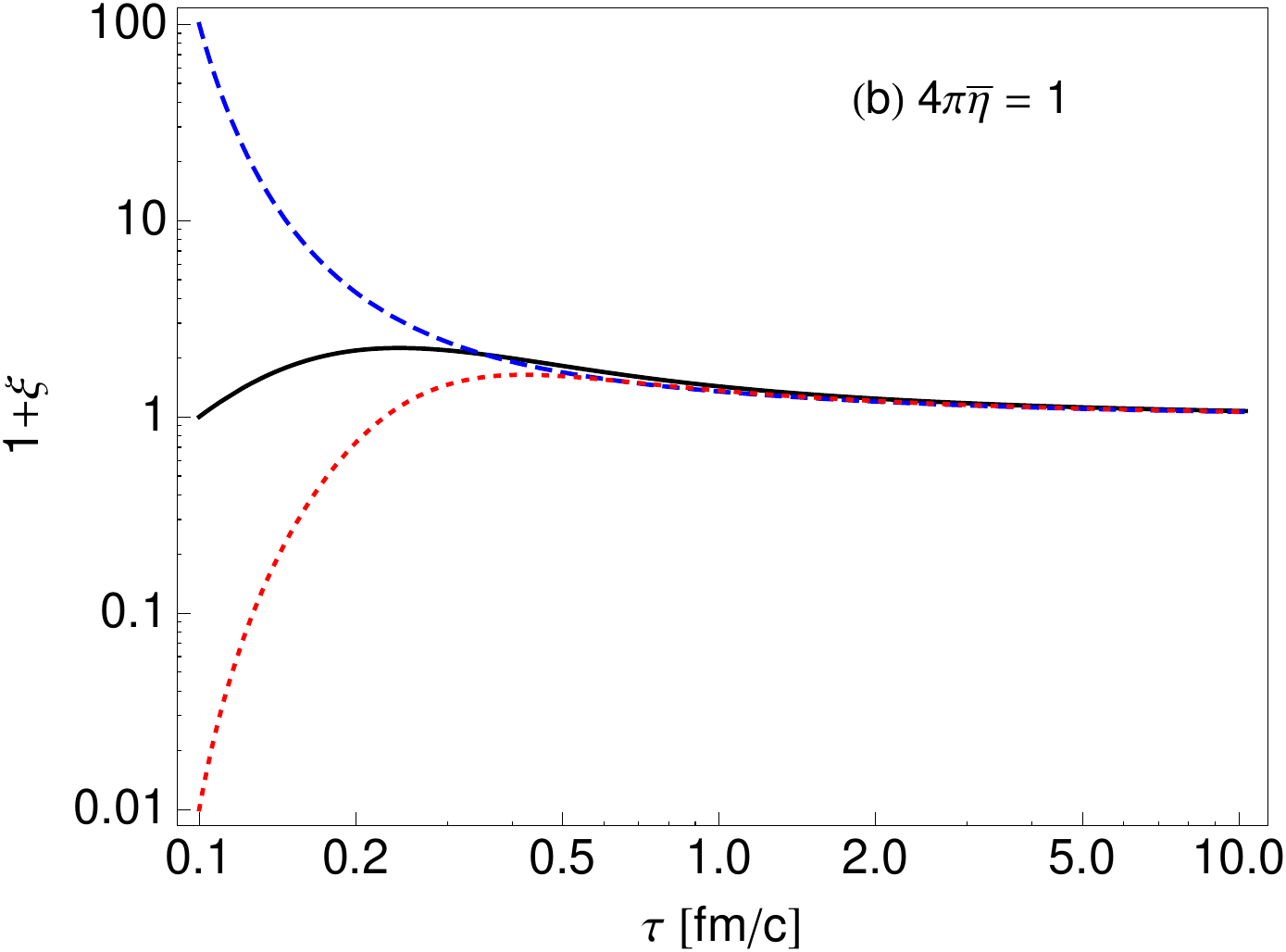}}\\
\subfigure{\includegraphics[angle=0,width=0.42\textwidth]{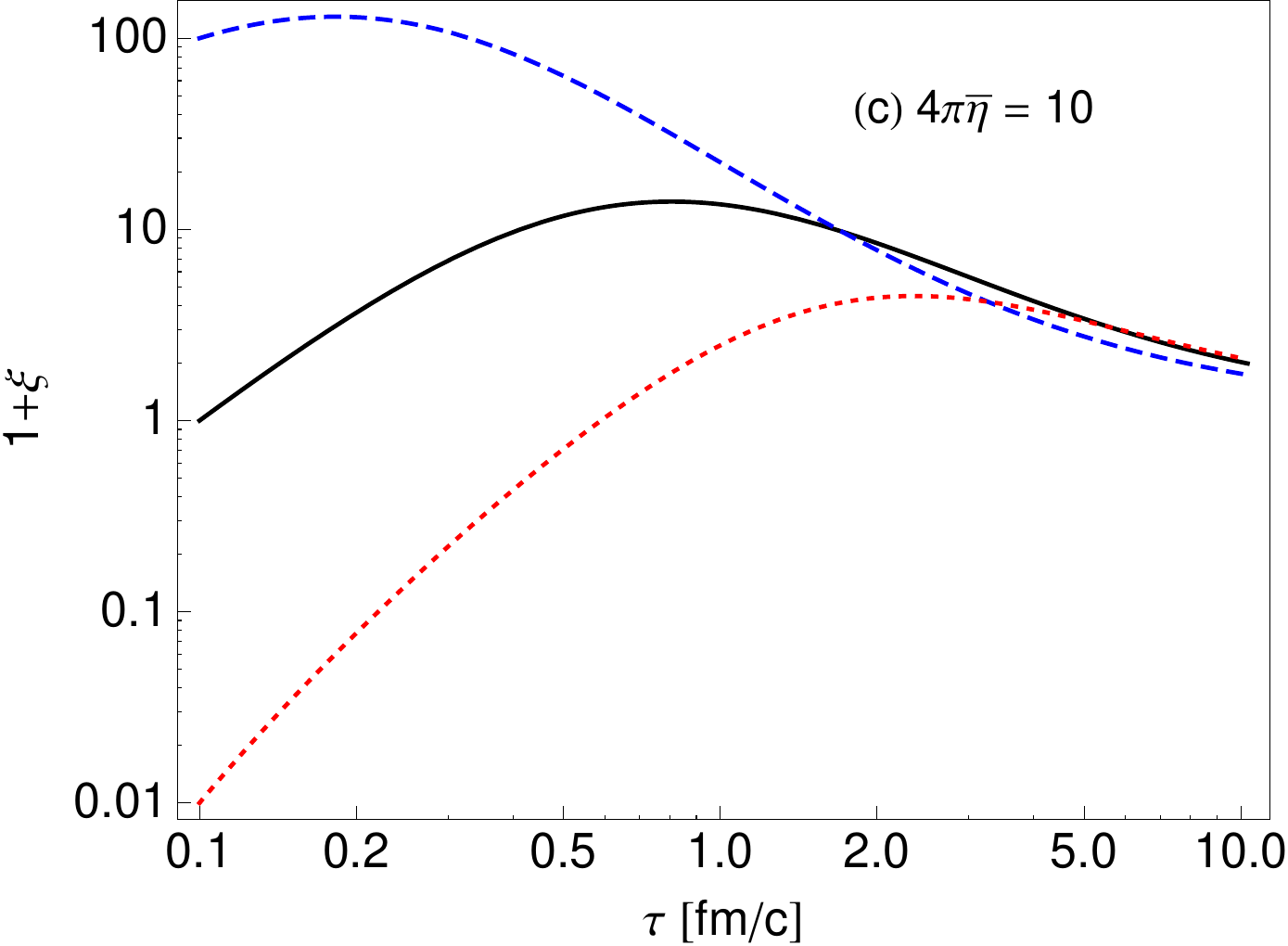}}
\end{center}
\caption{(Color online) Time dependence of the anisotropy parameter $\xi$ for (a) constant $\tau_{\rm eq}$, (b) time varying
$\tau_{\rm eq}$ with $4\pi\bar\eta = 1$, and (c)  time varying
$\tau_{\rm eq}$ with $4\pi\bar\eta = 10$.  In each plot we show three different initial
values of $\xi$:  $\xi_0=99$ (dashed line), $\xi_0=0$ (solid line), and $\xi_0=-0.99$ (dotted line).
}
\label{fig:plot_background}
\end{figure}

In Fig.~\ref{fig:plot_background} we show the time dependence of the anisotropy parameter $\xi$. The initial time of the 
hydrodynamic evolution is taken to be $\tau_0 = 0.1$~fm/c, and the final time is taken to be $\tau_f = 10$~fm/c. We consider 
three different choices of  the initial conditions corresponding to three different values of the initial anisotropy $\xi_0=\xi(\tau_0)$:
$\xi_0=99$ (dashed lines), $\xi_0=0$ (solid lines), and \mbox{$\xi_0=-0.99$} (dotted lines). The initial transverse-momentum scale 
$\Lambda_0$ is taken to be \mbox{$\Lambda_0 = $~(1 GeV)$\cdot (1+\xi_0)^{1/6}$} in each case. The factor of $(1+\xi_0)^{1/6}$ 
guarantees that the initial number density of the background is the same for all values of $\xi_0$ considered.
We show the proper-time evolution of the anisotropy parameter, $\xi$, in three different
cases:  (a) fixed $\tau_{\rm eq} =$~1~fm/c, (b) $\tau_{\rm eq}$ given by Eq.~(\ref{taueqdyn}) with $4\pi\bar\eta =1$,
and (c) Eq.~(\ref{taueqdyn}) with $4\pi\bar\eta =10$.   In the case of $\tau_{\rm eq}$ = 1 fm/c shown in 
Fig.~\ref{fig:plot_background}(a) the anisotropy vanishes at late times.  When $\tau_{\rm eq}$ is given by 
Eq.~(\ref{taueqdyn}) (see Fig.~\ref{fig:plot_background}(b,c)) a finite anisotropy remains at late times which is 
consistent with 2nd-order viscous hydrodynamics.  Note that the initial growth of the anisotropy is connected with the effects 
of free streaming which dominate the very early dynamics \cite{Florkowski:2010cf,Martinez:2010sc}. 

\subsection{First order}

At first order in fluctuations we obtain
\begin{eqnarray}
&& p^{\mu }\partial _{\mu } \delta Q^\pm_{i}(x,p)
\pm g {\mbox{\boldmath $\epsilon$}}_{i}\cdot {\bf F}^{\mu \nu }
p_{\nu } \partial_{\mu }^{p}  Q_0^\pm (x,p) \nonumber \\
&& \hspace{1.0cm} = - p^\mu U_\mu \frac{\delta Q^\pm_{i}(x,p)}{\tau_{\rm eq}},  
\label{kineq1}
\end{eqnarray}
\begin{eqnarray}
&& p^{\mu }\partial _{\mu } \delta G_{ij}(x,p) 
+g{\mbox{\boldmath $\eta$}}_{ij}\cdot {\bf F}^{\mu \nu } 
p_{\nu }\partial _{\mu }^{p}  G_0(x,p) \nonumber \\
&& \hspace{1.0cm} = - p^\mu U_\mu \frac{\delta G_{ij}(x,p)}{\tau_{\rm eq}}.  
\label{kineg1}
\end{eqnarray}

In the following we use the boost-invariant variables introduced in Refs.~\cite{Bialas:1984wv,Bialas:1984ap} 
\begin{equation}
w = tp_{\Vert }-zE,
\label{binvv1}
\end{equation}
and 
\begin{equation}
v = Et-p_{\Vert }\ z=\sqrt{w^{2}+m_{\perp }^{2} \tau^2}.  
\label{binvv2}
\end{equation}
From these two equations one can easily find the energy and the longitudinal momentum of a particle 
\begin{equation}
E=p^{0}=\frac{vt+wz}{\tau^2},\quad p_{\Vert }=\frac{wt+vz}{\tau^2}.  \label{binvv3}
\end{equation}
In addition, we have 
\begin{equation}
w=\tau m_{\perp }\sinh \left( y-\eta \right) ,\qquad v=\tau m_{\perp }\cosh
\left( y-\eta \right) .  \label{binvv4}
\end{equation}

Since the distribution functions are Lorentz scalars, they may depend only on $\tau$, $p_\perp$, and $w$. Therefore, we find the general boost-invariant form of the terms appearing in Eqs.~(\ref{kineq1}) and (\ref{kineg1})
\begin{eqnarray}
p^\mu U_\mu &=& \frac{v}{\tau}, \nonumber \\
p^{\mu }\partial _{\mu } \delta f &=& \frac{v}{\tau} \frac{\partial}{\partial \tau} \delta f, \label{binvder} \\
{\bf F}^{\mu \nu } p_{\nu } \partial_{\mu }^{p}  f_0 &=& 
{\mbox{\boldmath $\cal E$}}_z v \frac{\partial f_0}{\partial w}. \nonumber
\end{eqnarray}
The Romatschke-Strickland distribution function takes the form
\begin{eqnarray}
f_0(\tau,w) &=& \exp\left[-\frac{1}{\Lambda(\tau) \tau} \sqrt{p_\perp^2 \tau^2 + (1+\xi(\tau)) w^2} \, \right] .
\nonumber \\
&&
\label{RSformWV}
\end{eqnarray}
Using Eqs.~(\ref{binvder}) we find the following equations
\begin{eqnarray}
\frac{\partial}{\partial \tau} \delta Q^\pm_{i} &=& \mp g \tau 
{\mbox{\boldmath $\epsilon$}}_{i} \cdot {\mbox{\boldmath $\cal E$}}_z
\frac{\partial f_0}{\partial w} - \frac{\delta Q^\pm_{i}}{\tau_{\rm eq}} , \\
\frac{\partial}{\partial \tau} \delta G_{ij} &=& - g \tau 
{\mbox{\boldmath $\eta$}}_{ij} \cdot {\mbox{\boldmath $\cal E$}}_z
\frac{\partial f_0}{\partial w} - \frac{\delta G_{ij}}{\tau_{\rm eq}}.
\label{EQS1}
\end{eqnarray}
The formal solutions of (\ref{EQS1}) for constant or time-dependent $\tau_{\rm eq}$ may be expressed as integrals
\begin{eqnarray}
\hspace{-3mm} \delta Q^\pm_{i} = \mp g {\mbox{\boldmath $\epsilon$}}_{i} \cdot
\int\limits_{\tau_0}^\tau d\tau^\prime \, \tau^\prime 
D(\tau,\tau^\prime)
{\mbox{\boldmath $\cal E$}}_z(\tau^\prime ) 
\frac{\partial f_0}{\partial w} (\tau^\prime, w),  \label{SOL1q} \\
\hspace{-3mm} \delta G_{ij} = - g {\mbox{\boldmath $\eta$}}_{ij} \cdot
\int\limits_{\tau_0}^\tau d\tau^\prime \, \tau^\prime 
D(\tau,\tau^\prime)
{\mbox{\boldmath $\cal E$}}_z(\tau^\prime ) 
\frac{\partial f_0}{\partial w} (\tau^\prime, w), \label{SOL1g}
\end{eqnarray}
where we have introduced the damping function
\begin{equation}
D(\tau,\tau^\prime) = \exp\left[-\int\limits_{\tau^\prime}^\tau \frac{d\tau^{\prime \prime}}{\tau_{\rm eq}(\tau^{\prime \prime})} \right].
\end{equation}

\section{Maxwell equations}
\label{sect:Max}

In order to close our system of equations we have to couple the fluctuations via currents to the Maxwell equations
\begin{equation}
\partial_\mu \mathbf{F}^{\mu \nu } = \mathbf{j}^\nu,
\end{equation}
where the color current is given by the expression
\begin{equation}
\mathbf{j}^\nu = g \int dP p^\nu \left[ \sum_{i=1}^3 {\mbox{\boldmath $\epsilon$}}_{i}
\left(\delta Q^+_{i} - \delta Q^-_{i} \right) + \sum_{i,j=1}^3 {\mbox{\boldmath $\eta$}}_{ij} 
\delta G_{ij} \right].
\end{equation}
Using the formal solutions for the distribution functions at first order, we obtain
\begin{eqnarray}
&& \mathbf{j}^\nu = -g^2 \int\limits_{\tau_0}^\tau d\tau^\prime \, \tau^\prime 
D(\tau,\tau^\prime)
\int dP \, p^\nu 
\frac{\partial f_0}{\partial w} (\tau^\prime, w) 
 \\
&& \times \left[ 
\sum_{i=1}^3 2 \,{\mbox{\boldmath $\epsilon$}}_{i} ({\mbox{\boldmath $\epsilon$}}_{i} \cdot {\mbox{\boldmath $\cal E$}}_z(\tau^\prime ))  
+ \sum_{i,j=1}^3 {\mbox{\boldmath $\eta$}}_{ij} ({\mbox{\boldmath $\eta$}}_{ij} \cdot {\mbox{\boldmath $\cal E$}}_z(\tau^\prime ))  
\right]  .  \nonumber
\end{eqnarray}
Here we used the property $\delta Q^-_{i}(\tau,p_\perp,w) = \delta Q^+_{i}(\tau,p_\perp,-w)$ to eliminate the antiquark distribution function.

The invariant measure in momentum space is
\begin{equation}
dP = \nu_g d^2p_\perp {dp_\parallel \over p^0} = \nu_g d^2p_\perp {dw \over v},
\label{dP}
\end{equation}
where $\nu_g = \nu_{\rm sf}/(2\pi)^3$ and $\nu_{\rm sf}$ denotes the number of internal degrees of freedom connected with spin or flavor ($\nu_{\rm sf}=4$ for quarks and $\nu_{\rm sf}=2$ for gluons). Using Eqs.~(\ref{norm}) and (\ref{dP}) we find
\begin{eqnarray}
&& \mathbf{j}^\nu = -g^2 \int\limits_{\tau_0}^\tau d\tau^\prime \, \tau^\prime 
D(\tau,\tau^\prime)
\int d^2p_\perp \int\limits_{-\infty}^\infty {dw \,p^\nu \over (2 \pi)^3 \,v} 
 \nonumber
 \\
&& \times \frac{\partial f_0}{\partial w} (\tau^\prime, w) \left[ 
2 \cdot 4 \cdot \frac{1}{2} \, {\mbox{\boldmath $\cal E$}}_z(\tau^\prime ) 
+  2 \cdot 3 \cdot {\mbox{\boldmath $\cal E$}}_z(\tau^\prime ) 
\right].   \nonumber
\end{eqnarray}

For the Lorentz index $\nu=0$, we use the symmetry of the distribution function under the change $w \to -w$ and write
\begin{eqnarray}
\mathbf{j}^0 &=&  10 \,z\, g^2 \int\limits_{\tau_0}^\tau d\tau^\prime \, 
D(\tau,\tau^\prime)  {\mbox{\boldmath $\cal E$}}_z(\tau^\prime ) 
 \frac{  (1+\xi^\prime)}{\Lambda^\prime  \tau^2} \label{j03}  \\
&& \times 
\int \frac{d^2p_\perp}{(2 \pi)^3} \int\limits_{-\infty}^\infty {dw \, w^2 \over  v}
\frac{f_0^\prime }{ \sqrt{p_\perp^2 \tau^{\prime 2} + (1+\xi^\prime) w^2}}.   \nonumber
\end{eqnarray}
Here, the primes denote the dependence on $\tau^\prime$, for example, $\Lambda^\prime = \Lambda(\tau^\prime)$ and $f_0^\prime=f_0(\tau^\prime,p_\perp,w)$.

The zeroth component of the Maxwell equation gives
\begin{equation}
\partial_3 {\bf F}^{3 0} = -\frac{z}{\tau} 
\frac{d{\mbox{\boldmath $\cal E$}}_z(\tau) }{d\tau},
\label{Max3}
\end{equation}
hence, Eqs.~(\ref{j03}) and (\ref{Max3}) yield~\footnote{Because of boost-invariance we obtain the same result from the Maxwell equations with $\nu=3$.} 
\begin{eqnarray}
&& \frac{d{\mbox{\boldmath $\cal E$}}_z(\tau) }{d\tau} =  - \frac{10 g^2}{(2 \pi)^2 \tau} \int\limits_{\tau_0}^\tau d\tau^\prime \, 
D(\tau,\tau^\prime)  {\mbox{\boldmath $\cal E$}}_z(\tau^\prime ) 
 \frac{  (1+\xi^\prime)}{\Lambda^\prime} \nonumber  \\
&& \times 
\int\limits_0^\infty dp^2_\perp \int\limits_0^\infty {dw \, w^2 \over  v}
\frac{f_0^\prime }{ \sqrt{p_\perp^2 \tau^{\prime 2} + (1+\xi^\prime) w^2}}.   \label{FE1}
\end{eqnarray}
To proceed, we introduce new variables $\gamma$ and $\phi$ defined by the relations
\begin{eqnarray}
\gamma \cos\phi &=& \frac{p_\perp}{\Lambda^\prime} , \nonumber \\
\gamma \sin\phi &=& (1+\xi^\prime) \frac{w}{\Lambda^\prime \tau^\prime}.
\end{eqnarray}
The integration over $\gamma$ (from 0 to $\infty$) and $\phi$ (from 0 to $\pi/2$) can be performed analytically 
\cite{Bialas:1987cn}, and one obtains
\begin{eqnarray}
\frac{d{\mbox{\boldmath $\cal E$}}_z(\tau) }{d\tau} = && 
- \frac{\kappa g^2}{\tau} \int\limits_{\tau_0}^\tau d\tau^\prime \, 
D(\tau,\tau^\prime)  {\mbox{\boldmath $\cal E$}}_z(\tau^\prime ) 
 \nonumber \\
&& \times \tau^\prime  \Lambda^{\prime \,2}  J\left(\frac{\tau \sqrt{1+\xi^\prime} }{\tau^\prime}\right), \nonumber  \\
\label{FE2}
\end{eqnarray}
where, for a Boltzmann distribution, $\kappa=10/(3 \pi^2)$ and the function $J(\tau\sqrt{1+\xi^\prime}/\tau^\prime)$ is defined by the formula \cite{Bialas:1987cn}
\begin{eqnarray}
J\left(\frac{\tau  \sqrt{1+\xi^\prime}}{\tau^\prime}\right) &=&
\frac{3}{4} \frac{b+1}{b^{3/2}} \left[ \frac{\pi}{2} + \arcsin \left(\frac{b-1}{b+1} \right) \right]
-\frac{3}{2b}, \nonumber \\
b &=& \frac{\tau^2}{\tau^{\prime \,2}} (1+\xi^\prime)-1.
\end{eqnarray}

Eq.~(\ref{FE2}) determines the oscillations of a uniform chromoelectric field in an arbitrary 
time-evolving anisotropic background. It forms the 
basis of our numerical calculations presented below. Interestingly, the form of (\ref{FE2}) is very similar to that obtained in 
\cite{Bialas:1987cn}. Eq.~(\ref{FE2}) is reduced to Eq.~(4.8) in \cite{Bialas:1987cn}, if we set $\xi^\prime=0$ in the 
argument of $J$, $\Lambda^\prime$ is replaced by the temperature $T^\prime$, and the damping function $D$ is taken to 
be equal to unity. 
Note that since the calculations in \cite{Bialas:1987cn} were performed using the quantum statistical distribution functions, 
the factor $\kappa$ in Ref.~\cite{Bialas:1987cn} equals $(N_c + N_f/2)/9 = 4/9$ for $N_c=3$ and $N_f=2$. The calculations 
presented in this paper can be also performed using quantum statistical distribution functions and, in that case, one obtains 
$\kappa=4/9$.  The results presented in Sec.~\ref{sect:results} use this choice of $\kappa$.

\subsection{Special Case: Longitudinal free streaming limit}

For the case of longitudinal free streaming it is possible to write the integro-differential equation (\ref{FE2}) as an ordinary 
differential equation.  In the limit $\tau_{\rm eq} \rightarrow \infty$ one has $\xi(\tau) = (\tau/\tau_{\rm iso})^2 - 1$ and 
$\Lambda(\tau) = \Lambda_0$ with $\tau_{\rm iso}=\tau_0/\sqrt{1+\xi(\tau_0)}$ being the point in time when $\xi=0$.  
Inserting these relations into (\ref{FE2}) gives
\begin{equation}
\frac{\tau}{J\!\left(\frac{\tau}{\tau_{\rm iso}}\right)}\frac{d{\mbox{\boldmath $\cal E$}}_z(\tau) }{d\tau} = 
- \omega_{\rm pl}^2 \int\limits_{\tau_0}^\tau d\tau^\prime \, 
{\mbox{\boldmath $\cal E$}}_z(\tau^\prime ) \tau^\prime \, ,
\label{FE2-fs}
\end{equation}
where $\omega_{\rm pl}^2 = \kappa g^2 \Lambda_0^2$.
Taking a derivative of both sides of this equation with respect to $\tau$ gives an ordinary differential equation
\begin{equation}
\frac{1}{\tau}\frac{d}{d\tau}\left(\frac{\tau}{J\!\left(\frac{\tau}{\tau_{\rm iso}}\right)}\frac{d{\mbox{\boldmath $\cal E$}}_z(\tau) }{d\tau}\right) = - \omega_{\rm pl}^2 {\mbox{\boldmath $\cal E$}}_z(\tau ) \, .
\label{FE2-fsde}
\end{equation}
which is supplemented by the initial conditions ${\mbox{\boldmath $\cal E$}}_z(\tau_0) = 
{\mbox{\boldmath $\cal E$}}_{z,0}$ and ${\mbox{\boldmath $\cal E$}}_z^\prime(\tau_0) = 0$, where the latter
condition follows from (\ref{FE2-fs}) upon setting $\tau=\tau_0$.

The differential equation above is nonlinear and must be solved numerically; however, at late times we can find
the asymptotic form of the solution by using
\begin{equation}
\lim_{\tau \rightarrow \infty} J\!\left(\frac{\tau}{\tau_{\rm iso}}\right) = \frac{3 \pi \tau_{\rm iso}}{4 \tau} 
+ {\cal O}\left(\left(\frac{\tau_{\rm iso}}{\tau}\right)^2\right) ,
\end{equation}
to obtain
\begin{equation}
\frac{1}{\tau}\frac{d}{d\tau}\left(\tau^2 \frac{d{\mbox{\boldmath $\cal E$}}_z(\tau) }{d\tau}\right) = - \mu {\mbox{\boldmath $\cal E$}}_z(\tau ) \, ,
\end{equation}
where $\mu \equiv 3 \pi \tau_{\rm iso} \omega_{\rm pl}^2/4$.  This differential equation has a solution of the form
\begin{equation}
\lim_{\tau \rightarrow \infty} \mbox{\boldmath $\cal E$}_z = \frac{1}{\sqrt{\mu\tau}} \left[ {\bf A} J_1\left(2 \sqrt{\mu \tau}\right) + {\bf B} Y_1\left(2 \sqrt{\mu \tau} \right) \right] ,
\label{FE2-asymp}
\end{equation}
where $J_1$ and $Y_1$ are Bessel functions of the first and second kind, respectively, and $\bf A$ and $\bf B$
collect undetermined constants that will be matched below.  Note that this agrees with
the result first obtained in App.~A subsection 2 of Ref.~\cite{Rebhan:2009ku}.\footnote{Their result is expressed 
in terms of the longitudinal vector potential.  One must compute the longitudinal electric field using 
$E^\eta = \Pi^\eta =\tau^{-1} \partial_\tau A_\eta$ in order to compare with our result.}

\section{Results}
\label{sect:results}

\begin{figure}[th!]
\begin{center}
\subfigure{\includegraphics[angle=0,width=0.42\textwidth]{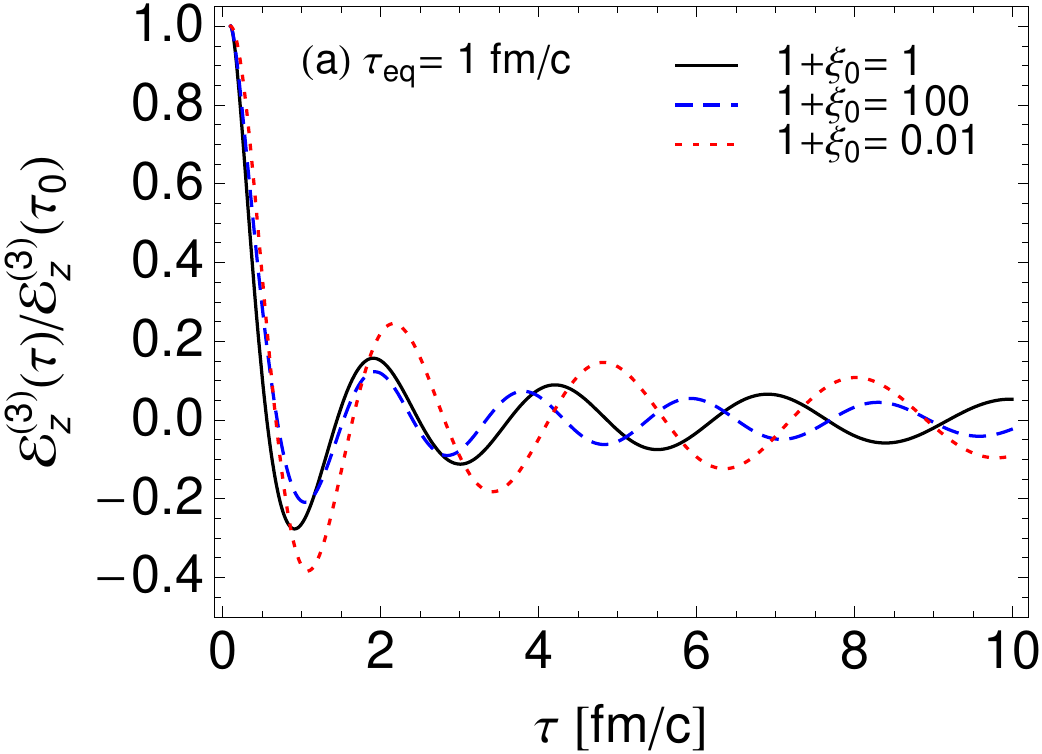}}\\
\subfigure{\includegraphics[angle=0,width=0.42\textwidth]{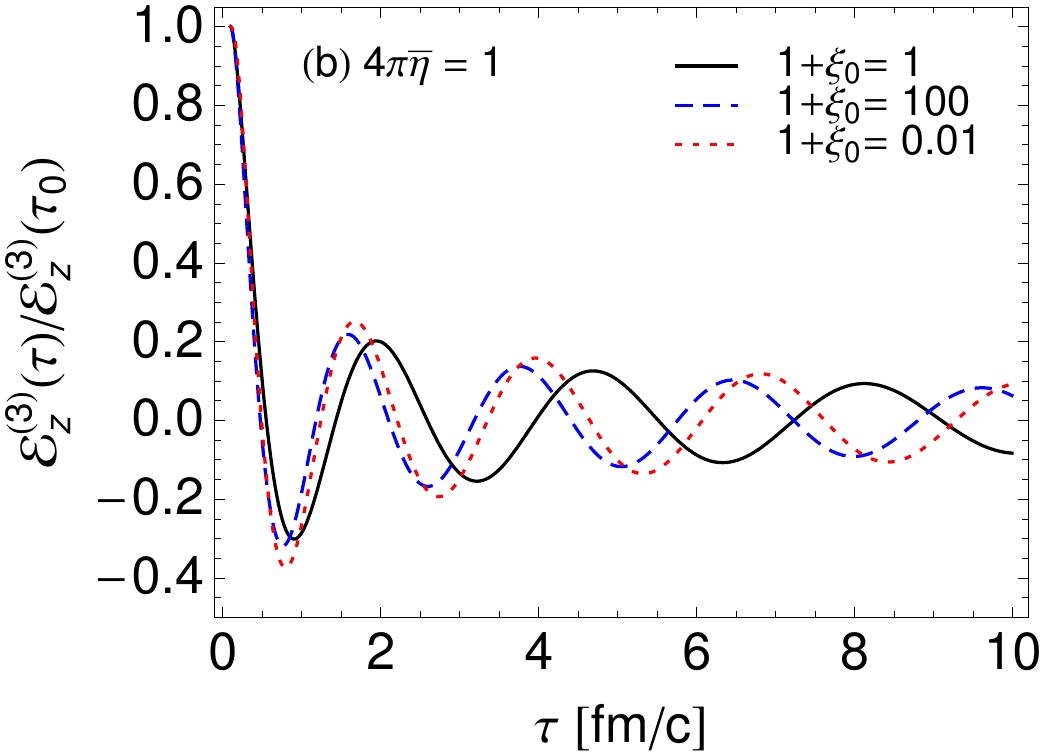}}\\
\subfigure{\includegraphics[angle=0,width=0.42\textwidth]{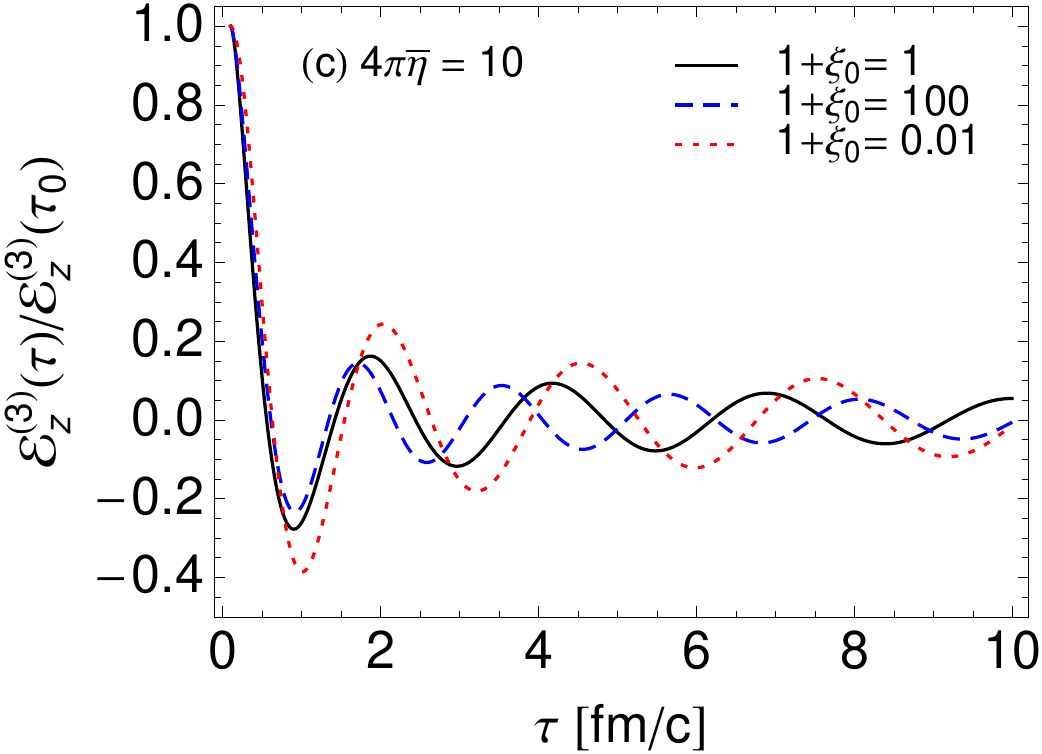}}
\end{center}
\caption{(Color online) Time dependence normalized of the normalized longitudinal chromoelectric field 
$\mbox{\boldmath $\cal E$}_z$ for (a) constant $\tau_{\rm eq}$, (b) time varying $\tau_{\rm eq}$ with $4\pi\bar\eta = 1$, 
and (c)  time varying
$\tau_{\rm eq}$ with $4\pi\bar\eta = 10$.  We turn off the damping manually by setting the damping function
$D(\tau,\tau^\prime) = 1$.  In each plot we show three different initial
values of $\xi$:  $\xi_0=99$ (dashed line), $\xi_0=0$ (solid line), and $\xi_0=-0.99$ (dotted line).
}
\label{fig:plot_ez_nodamp}
\end{figure}

\begin{figure}[th!]
\begin{center}
\subfigure{\includegraphics[angle=0,width=0.42\textwidth]{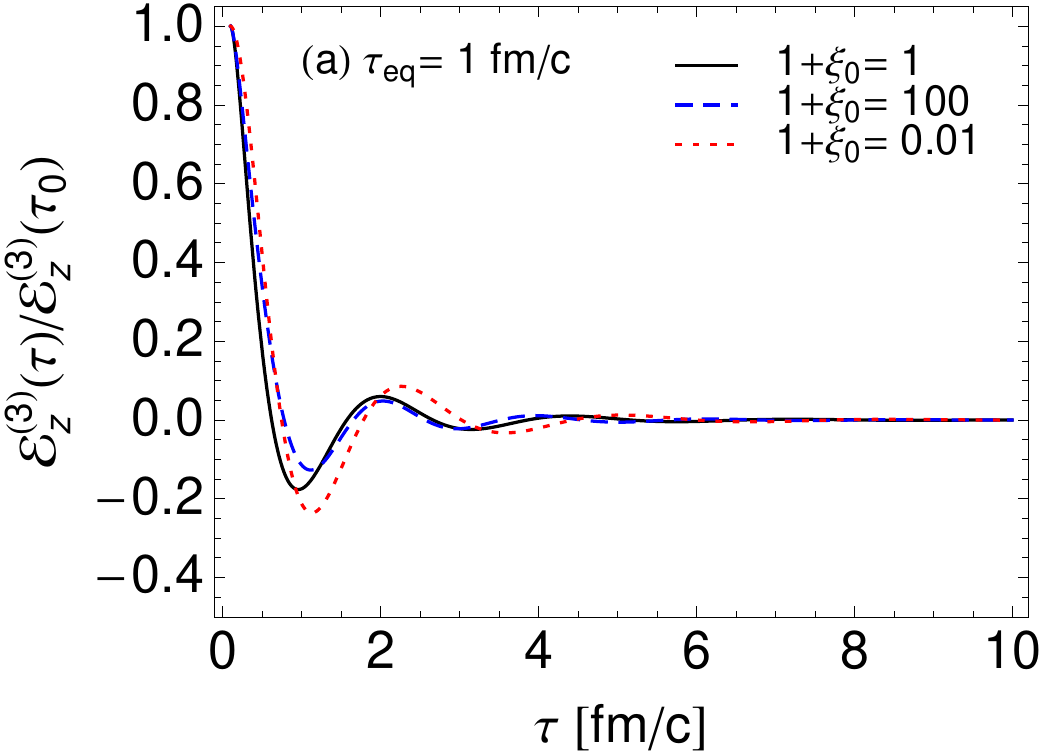}}\\
\subfigure{\includegraphics[angle=0,width=0.42\textwidth]{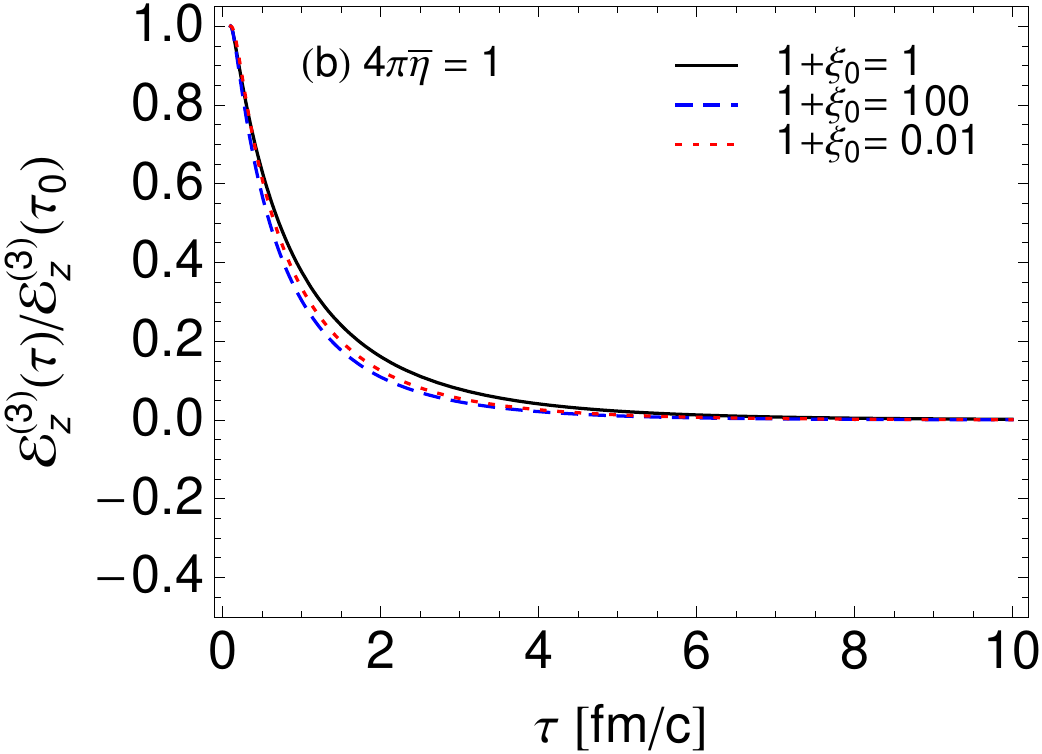}}\\
\subfigure{\includegraphics[angle=0,width=0.42\textwidth]{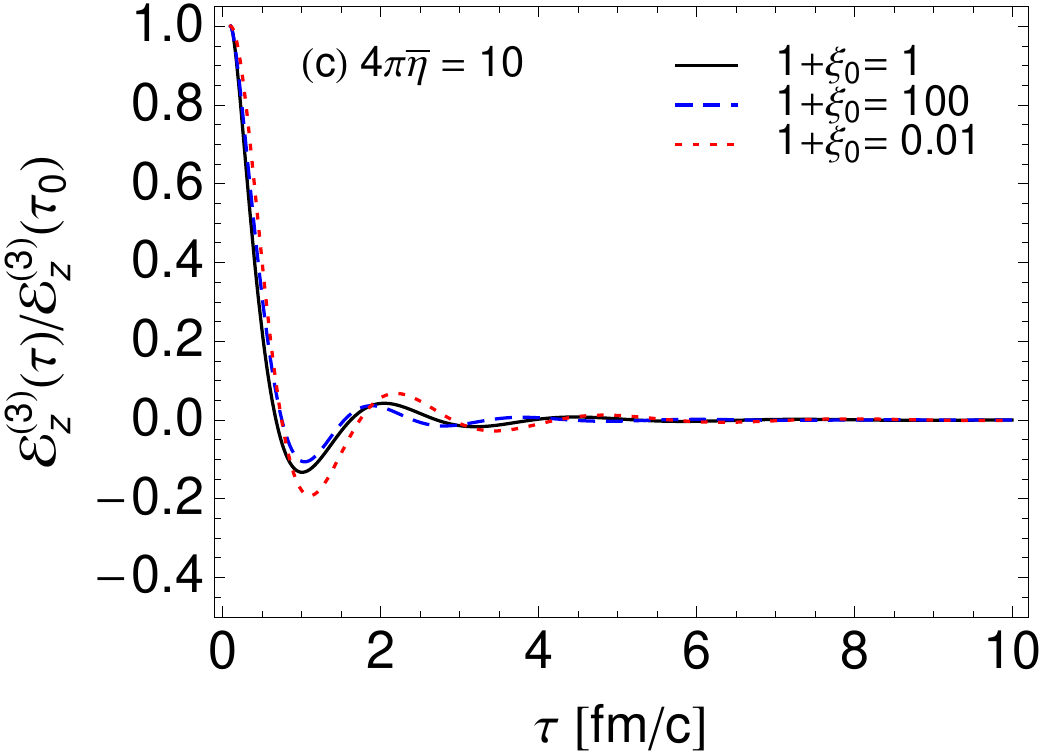}}
\end{center}
\caption{(Color online) Time dependence of the normalized longitudinal chromoelectric field $\mbox{\boldmath $\cal E$}_z$ 
for (a) constant $\tau_{\rm eq}$, (b) time varying $\tau_{\rm eq}$ with $4\pi\bar\eta = 1$, and (c)  time varying
$\tau_{\rm eq}$ with $4\pi\bar\eta = 10$.  In each plot we show three different initial
values of $\xi$:  $\xi_0=99$ (dashed line), $\xi_0=0$ (solid line), and $\xi_0=-0.99$ (dotted line).
}
\label{fig:plot_ez_damp}
\end{figure}

For the numerical results we choose a particular direction of the chromolectric field by aligning it initially in the `3'
direction, i.e. $\mbox{\boldmath $\cal E$}_z(\tau_0) = ({\cal E}_z^{(3)}(\tau_0),0)$.  The evolution keeps 
$\mbox{\boldmath $\cal E$}_z$ in the `3' direction at all times.
In Fig.~\ref{fig:plot_ez_nodamp} we plot the time dependence of the normalized longitudinal chromoelectric field 
${\cal E}_z^{(3)}$ for (a) constant $\tau_{\rm eq}$, (b) time varying $\tau_{\rm eq}$ with $4\pi\bar\eta = 1$, 
and (c)  time varying $\tau_{\rm eq}$ with $4\pi\bar\eta = 10$.  We turn off the damping manually by setting the damping 
function $D(\tau,\tau^\prime) = 1$.  In each plot we show three different initial values of $\xi$:  $\xi_0=99$ (dashed line), 
$\xi_0=0$ (solid line), and $\xi_0=-0.99$ (dotted line).  For each $\xi_0$ we have adjusted $\Lambda_0$ as described
in Sec.~\ref{subsect:background} in order to guarantee that the initial number density is held constant.  This figure 
demonstrates that, although we have varied our assumed value of $\xi_0$ over a large range corresponding to initially 
extremely prolate ($\xi_0 = -0.99$) to extremely oblate ($\xi_0 = 99$), if the results are normalized such that the initial 
number densities are held constant, the resulting oscillations are not dramatically different.  This is due to the fact that what 
sets the time scale for the plasma oscillations is $\omega_{\rm pl}$ and, generally, one has $\omega_{\rm pl} 
\sim n(\tau_0)/\Lambda_0$.  However, despite being qualitatively similar, there are important quantitative differences which 
remain.

In Fig.~\ref{fig:plot_ez_damp} we plot the time dependence of the normalized longitudinal chromoelectric field 
$\mbox{\boldmath $\cal E$}_z$ for (a) constant $\tau_{\rm eq}$, (b) time varying $\tau_{\rm eq}$ with $4\pi\bar\eta = 1$, 
and (c)  time varying $\tau_{\rm eq}$ with $4\pi\bar\eta = 10$.  We now include the damping function in the integrand
of the integro-differential equation.  In each plot we show three different initial values of $\xi$:  $\xi_0=99$ (dashed line), 
$\xi_0=0$ (solid line), and $\xi_0=-0.99$ (dotted line).  Once again, for each $\xi_0$ we have adjusted $\Lambda_0$ as 
described in Sec.~\ref{subsect:background} in order to guarantee that the initial number density is held constant.   From 
this figure we see that the damping function $D(\tau,\tau^\prime)$ has an extremely important impact on the time 
evolution of the oscillations of uniform longitudinal chromoelectric fields.\footnote{We made a preliminary study
of the impact of collisional damping on unstable modes and found that the damping serves only to slightly
reduce the growth rate of unstable modes.  Results from this study will be reported elsewhere.}

\begin{figure}[t]
\begin{center}
\includegraphics[angle=0,width=0.45\textwidth]{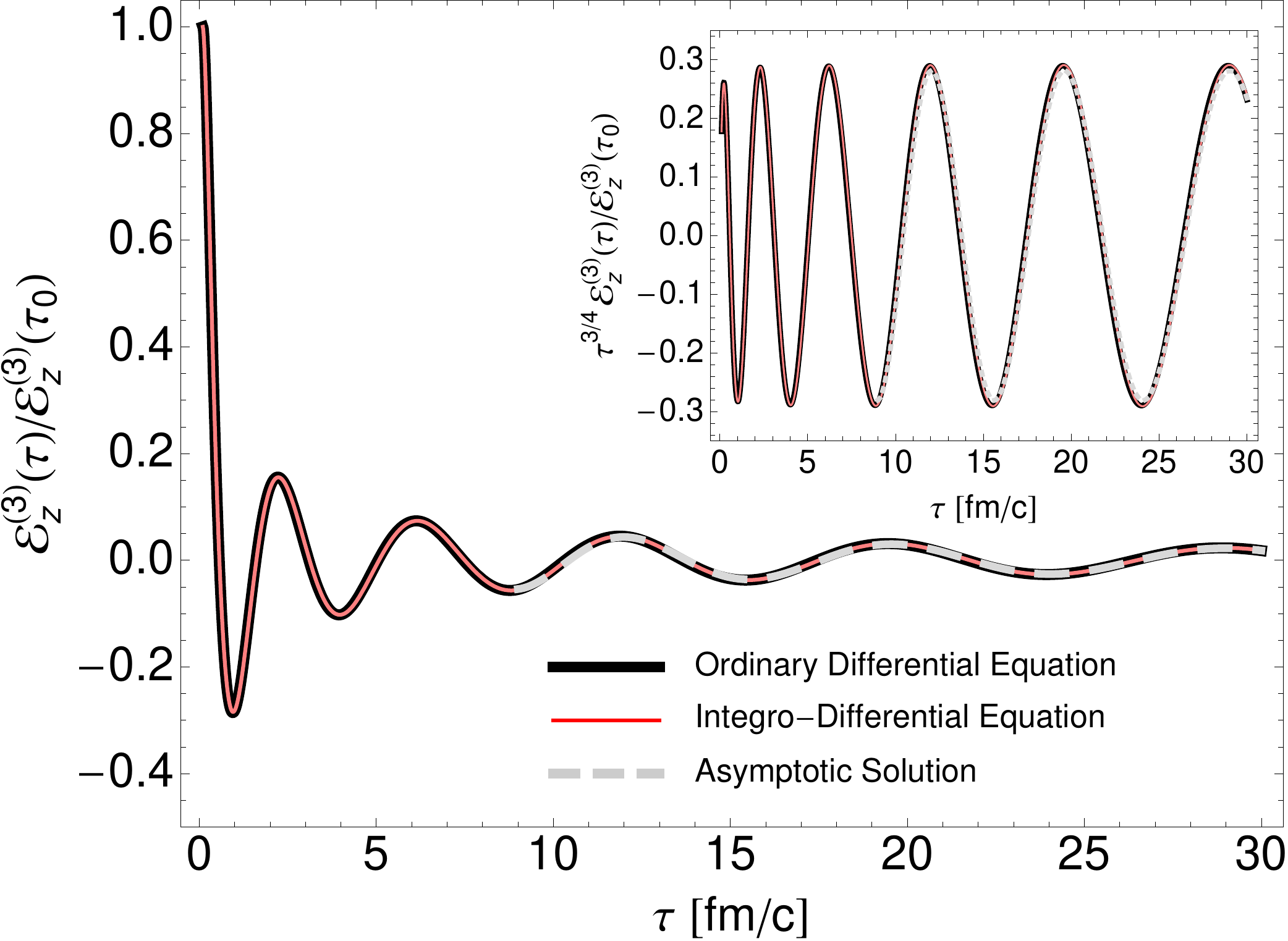}
\end{center}
\caption{(Color online) Comparison of solution to the ordinary differential equation (\ref{FE2-fsde}) (thick black line), the 
integro-differential equation (\ref{FE2}) (thin red line), and the asymptotic solution (\ref{FE2-asymp}) (thick gray dashed line) 
in the case of longitudinal free streaming ($\tau_{\rm eq}=\infty$).  The inset shows the three results scaled by $\tau^{3/4}$.
}
\label{fig:plot_fs_comp}
\end{figure}

In Fig.~\ref{fig:plot_fs_comp} we compare solutions to the ordinary differential equation (\ref{FE2-fsde}) (thick black line), the 
integro-differential equation (\ref{FE2}) (thin red line), and the asymptotic solution (\ref{FE2-asymp}) (thick gray 
dashed line) in the case of longitudinal free streaming ($\tau_{\rm eq}=\infty$).  The inset shows the three results 
scaled by $\tau^{3/4}$ in order to make a more precise comparison of the late time behavior.  For the asymptotic solution 
the unknown constants appearing in (\ref{FE2-asymp}) were fixed by matching numerically to the full solution at 
$\tau = 10^4$ fm/c.  As can be seen from this figure, our numerical solution to the integro-differential equation 
(\ref{FE2})  gives the same result as direct solution of the differential equation (\ref{FE2-fsde}) and at late times both 
agree well with the asymptotic solution.  This gives us confidence that the numerical method used for solution of the 
integro-differential equation (\ref{FE2}), in the general case, is reliable.

\section{Conclusions}
\label{sect:concl}

In this paper we studied the oscillations of a uniform longitudinal chromoelectric field in a dynamically-evolving 
momentum-space anisotropic background.  The required anisotropic hydrodynamics equations were
obtained from the first two moments on the Boltzmann-Vlasov equation using a spheroidal form for the local
rest frame one-particle distribution function.  The resulting anisotropic hydrodynamics equations provided the proper-time 
dependence of the local transverse momentum scale, $\Lambda(\tau)$, and momentum-space anisotropy, $\xi(\tau)$.
We then expanded the Boltzmann-Vlasov equation to first order in fluctuations and coupled these fluctuations
to the Maxwell equations.  From this procedure we obtained an integro-differential equation which governs
the time evolution of a uniform longitudinal chromoelectric field.  The integro-differential equation allows for
an arbitrary time dependence of the scale $\Lambda(\tau)$ and momentum-space anisotropy $\xi(\tau)$ provided
that the system is boost invariant and transversely homogeneous.  The integro-differential equation obtained
also includes the effect of collisional damping of the oscillation.

Having obtained the integro-differential equation necessary, we proceeded to solve it numerically for a variety of 
different initial momentum-space anisotropies using two different assumptions for the relaxation time 
$\tau_{\rm eq}$.  We considered (i) the case of constant $\tau_{\rm eq}$, in which case the late-time dynamics is 
that of ideal hydrodynamics, and (ii) a time-dependent $\tau_{\rm eq}$ that is inversely proportional to the local
inverse average transverse momentum, in which case the late-time dynamics is that of 2nd-order viscous
hydrodynamics.  We showed that for fixed initial number density the effect of time-varying momentum-space
anisotropy is important but not overwhelming large.  However, we found the effect of collisional damping to be
quite important for the dynamics of stable chromoelectric oscillations.

We should stress that the results presented here are exploratory in the sense that we have only studied the 
dynamics of a stable uniform chromoelectric field; however, the general method used here could have a wide-ranging
application in the study of the dynamics of all stable and unstable modes of a longitudinally expanding QGP.  Previous 
studies in this direction have been restricted to the limiting case of a longitudinally free-streaming anisotropic background 
\cite{Romatschke:2006wg,Rebhan:2008uj,Rebhan:2009ku}.  We were able to check our results
against the free-streaming results obtained by Rebhan and Steineder \cite{Rebhan:2009ku} and found that
we are able to reproduce their results for the dynamics of a uniform longitudinal chromoelectric field.  In addition, 
we were able to express the free-streaming evolution of a uniform chromoelectric field as an ordinary differential 
equation, albeit a highly-nonlinear one.  It would be very interesting to see if the results obtained here could be used to 
obtain modified hard-loop equations of motion that would allow one to simulate the full non-Abelian 
dynamics of stable and unstable modes in a realistically time-evolving anisotropic background.  Work along these lines 
is in progress.

\begin{acknowledgments}

W.F. and R.R. were supported by the Polish Ministry of Science and Higher Education under Grant No.~N N202 263438.                  
M.S. was supported by NSF grant No.~PHY-1068765 and the Helmholtz International Center for FAIR LOEWE program.

\end{acknowledgments}

\bibliography{osc_1}

\end{document}